\documentclass[10pt,twocolumn]{IEEEtran}

\usepackage{gensymb}
\usepackage{amsmath}
\usepackage{amsfonts}
\usepackage{epsfig}
\usepackage{amssymb}
\usepackage{cite}
\usepackage{hhline}
\usepackage{multirow}
\usepackage{framed}
\usepackage{xcolor}
\usepackage{lipsum}
\usepackage{bm}
\usepackage{url}

\usepackage{makecell}

\usepackage{color}
\usepackage{graphicx}
\usepackage{subfigure}

\usepackage{tikz}
\usepackage{pgfplots}

\newcommand{\NEW}[1]{#1}

\begin{document}

\title{Design Aspects of Short Range Millimeter Wave Networks: A MAC Layer Perspective}

\author{Hossein Shokri-Ghadikolaei,~\IEEEmembership{Student Member,~IEEE,} Carlo Fischione,~\IEEEmembership{Member,~IEEE,} \newline Petar Popovski,~\IEEEmembership{Senior Member,~IEEE,} and Michele Zorzi,~\IEEEmembership{Fellow,~IEEE} %
\thanks{H. Shokri-Ghadikolaei and C. Fischione are with KTH Royal Institute of Technology, Stockholm,
Sweden (email: \{hshokri, carlofi\}@kth.se).}
\thanks{P.~Popovski is with the Department of Electronic Systems, Aalborg University, Aalborg, Denmark (e-mail: petarp@es.aau.dk).}
\thanks{M.~Zorzi is with the Department of Information Engineering, University of Padova, Padova, Italy (e-mail: zorzi@dei.unipd.it).}
}

\maketitle

\begin{abstract}
Increased density of wireless devices, ever growing demands for extremely high data rate, and spectrum scarcity at microwave bands make the millimeter wave (mmWave) frequencies an important player in future wireless networks. However, mmWave communication systems exhibit severe attenuation, blockage, deafness, and may need microwave networks for coordination and fall-back support. To compensate for high attenuation, mmWave systems exploit highly directional operation, which in turn substantially reduces the interference footprint. The significant differences between mmWave networks and legacy communication technologies challenge the classical design approaches, especially at the medium access control (MAC) layer, which has received comparatively less attention than PHY and propagation issues in the literature so far.
In this paper, the MAC layer design aspects of short range mmWave networks are discussed. \NEW{In particular, we explain why current mmWave standards fail to fully exploit the potential advantages of short range mmWave technology, and argue for the necessity of new collision-aware hybrid resource allocation frameworks with on-demand control messages, the advantages of a collision notification message, and the potential of multihop communication to provide reliable mmWave connections.}
\end{abstract}

\section{Introduction}\label{sec: introduction}
\NEW{
Millimeter wave (mmWave) wireless communications are one of the most promising candidates to support extremely high data rates in future wireless networks~\cite{rappaport2014mmWaveBook,Niu2015Survey,Nitsche2014IEEE}. 
MmWave communications are attractive for many applications such as ultra short range communications, augmented reality, data centers, vehicular networks, mobile offloading, mobile fronthauling, and in-band backhauling. Due to their great commercial potential, several international activities have emerged to standardize mmWave communications in wireless personal and local area networks (WPANs and WLANs). Examples include IEEE~802.15.3c, ECMA~387~\cite{rappaport2014mmWaveBook}, IEEE~802.11ad~\cite{Nitsche2014IEEE}, WirelessHD, WiGig, and recently the IEEE~802.11ay study group on next generation 60~GHz.\footnote{Detailed information about these projects can be found at the following addresses: \url{http://www.wirelesshd.org} (WirelessHD), \url{http://wirelessgigabitalliance.org} (WiGig), and \url{http://www.ieee802.org/11/Reports/ng60_update.htm} (IEEE~802.11ay), respectively.}
}

Special propagation features and hardware constraints of mmWave systems introduce many new challenges in the design of efficient physical, medium access control (MAC), and routing layers.
The severe channel attenuation, vulnerability to obstacles, directionality of mmWave communications, the reduced interference footprint, and high signaling overhead demand a thorough reconsideration of traditional protocols and design principles, especially at the MAC layer.

\NEW{In this paper, we focus on short range mmWave networks. Compared to~\cite{rappaport2014mmWaveBook,Nitsche2014IEEE,Niu2015Survey} that survey either the existing standards or the research literature, we deliver original contributions based on the features specific to mmWave networks that are mostly ignored in the design of the existing mmWave standards.
To distinguish this paper from~\cite{shokri2015mmWavecellular} that discusses MAC layer design for mmWave cellular networks, we should notice the following important differences, which are more relevant to our studies, between short range and cellular networks: (i) short range networks may rely on carrier sensing among terminals,
and (ii) they may use multihop communications, which may also affect traffic patterns.
In this paper, we show that, contrary to mainstream belief, a mmWave network may exhibit both noise-limited and interference-limited regimes. We highlight the significant mismatch between transmission rates of control and data messages, which challenges the MAC layer efficacy of the existing mmWave standards in dense deployment scenarios. We also raise the prolonged backoff time problem and discuss the beam training overhead and its consequences such as the alignment-throughput tradeoff. To address these new problems, we discuss the necessity of new collision-aware hybrid resource allocation protocols that facilitate concurrent transmissions with QoS guarantees, and also the need for a more efficient retransmission policy. We argue the benefits of a hybrid reactive/proactive control plane to minimize the signaling overhead and propose, for this purpose, a new MAC layer message, which is also able to alleviate the prolonged backoff time. Finally, we discuss the potential of multihop communication techniques to compensate for the error-prone mmWave physical layer, provide reliable mmWave connections, and extend mmWave communication range.}

\NEW{Throughout this paper, we identify critical MAC layer aspects of the existing mmWave standards that may limit the efficacy and use cases of future mmWave networks. The detailed discussions and proposed solution approaches presented in this paper provide useful insights for future and emerging mmWave network technologies, such as IEEE~802.11ay.}

The rest of this paper is organized as follows. In Section~\ref{sec: fundamentals}, we describe the essential aspects of mmWave networks. In Section~\ref{sec: standardization}, existing mmWave standards are briefly reviewed. \NEW{Section~\ref{sec: Gap-Analysis} presents new fundamental aspects that are missing in the current standards, followed by MAC design guidelines in Section~\ref{sec: MAC-Design-Aspects}.} Concluding remarks are presented in Section~\ref{sec: concluding-remarks}.

\section{Fundamentals}\label{sec: fundamentals}

\subsection{The Directed mmWave Wireless Channel}\label{subsec: mmWave-channel}
MmWave communications use frequencies in the range 30--300~GHz, though the frequencies 6--30~GHz are also often referred to as mmWave\cite{shokri2015mmWavecellular}. The main characteristics of mmWave systems are high path-loss, large bandwidth, short wavelength/high frequency, and high penetration loss. Very small wavelengths allow the implementation of massive numbers of antenna elements in the current size of radio chips, which boosts the achievable antenna gain at the cost of extra signal processing. Such a gain can largely or even completely compensate for the higher path-loss of mmWave systems without any extra transmission power. Moreover, directional communications introduce the concept of directional spatial channel, i.e., a channel can be established in a specific direction with a range that varies according to the directionality level.

\subsection{Beam Training}\label{sec: beamsearching}
The use of low-complexity and low-power mmWave devices, along with the massive number of antennas, make traditional digital beamforming based on instantaneous channel state information very expensive. Instead, the existing standards establish a mmWave link using analog beamforming (also called beam-searching) based on pre-defined beam steering vectors (beam training codebook), each covering a certain direction with a certain beamwidth~\cite{rappaport2014mmWaveBook,Niu2015Survey,Nitsche2014IEEE}.
Current standards suggest a three-stage beam-searching technique to reduce alignment overhead. After a quasi-omnidirectional (low resolution pattern) sweep, a coarse grained sector-level sweep (second level resolution pattern) is performed, followed by a beam-level refinement phase (the highest resolution pattern specified in the codebook). An exhaustive search over all possible transmission and reception directions is applied in each level through a sequence of pilot transmissions.  The combination of vectors that maximizes the signal-to-noise ratio (SNR) is then selected for the beamforming.
\NEW{IEEE~802.11ad allows investigation of multiple beamforming vectors within a message, as opposed to sending separate training on each beamforming vector, which is the approach adopted in IEEE~802.15.3c. This
modification makes it possible to explore multiple beam patterns with lower overall overhead in IEEE~802.11ad. One of the main drawbacks of analog beamforming is the lack of multiplexing gain, which is addressed by the hybrid digital/analog beamforming architecture, see~\cite[Section~II-C]{shokri2015mmWavecellular} for more details.
}

\subsection{Deafness and Blockage}
Directional communications and vulnerability to obstacles in mmWave networks have two main consequences~\cite{rappaport2014mmWaveBook}: (1) deafness and (2) blockage.
\emph{Deafness} refers to the situation in which the main beams of the transmitter and the receiver do not point to each other, preventing the establishment of a directional communication link. \NEW{Deafness introduces a time consuming procedure for beam searching or \emph{alignment}; an operation in which two beams are pointing to each other, so that the link budget between the transmitter and the receiver is maximized.} The alignment procedure complicates the link establishment phase; however, it substantially reduces multiuser interference~\cite{Singh2011Interference}, as the receiver listens only to a specific directed channel.
In the extreme case, multiuser interference is almost completely suppressed and no longer limits the throughput, so that a mmWave network may operate in a noise-limited regime, unlike conventional interference-limited networks.\footnote{\NEW{Not being in an interference-limited regime does not necessarily imply that a network operates in a noise-limited regime, rather it only implies that the throughput per channel use is limited by the noise power. The network throughput performance, however, can be limited by other factors such as the signaling overhead, as will be argued in Section~\ref{sec: rate-mismatch}.}} This unique feature makes mmWave suitable for very dense deployments of infrastructure nodes and terminals.
\emph{Blockage} instead refers to very high attenuation due to obstacles (e.g., 35~dB due to the human body~\cite{shokri2015mmWavecellular}) that cannot be solved by just increasing the transmission power or increasing the antenna gain using narrower beams. \NEW{Overcoming blockage requires a search for alternative directed mmWave channels that are not blocked, which however entails a new alignment procedure and the consequent overhead.}


\subsection{Control Channel}
Many operations such as establishing a communication channel, discovering neighbors, exchanging routing information, and coordinating channel access rely on the exchange of signaling messages on a control channel. The characteristics of mmWave communications introduce fall-back and directionality tradeoffs, which also appear in mmWave cellular networks~\cite{shokri2015mmWavecellular}. The \emph{fall-back} tradeoff is the tradeoff between sending control messages through a mmWave or a microwave channel. The mmWave channel is subject to blockage, reducing the reliability of the control channel. A dedicated microwave control channel facilitates network synchronization and broadcasting at the expense of higher hardware complexity and energy consumption, since an extra transceiver should be tuned on the microwave control channel~\cite{nitsche2015steering}. \NEW{Moreover, a microwave control channel cannot be used to estimate the mmWave channel and adopt proper beamforming. This is a serious drawback that may hinder the use of hybrid beamforming in future mmWave networks.}
The \emph{directionality} tradeoff refers to the option of establishing a control channel in omnidirectional or directional operation modes. An omnidirectional control channel alleviates the deafness problem at the expense of being subject to a very short range, whereas a directional one increases the coverage with extra alignment overhead.
Altogether, we may have two justifiable control channels: (1) omnidirectional-microwave, which is employed in ECMA~387, and (2) directional-mmWave,\footnote{Note that realizing a control channel in the mmWave band with omnidirectional transmission and/or reception while having antenna gains for data transmission introduces a mismatch between the ranges at which a link with reasonable data rate can be established and the range at which control messages can be exchanged. Such a mismatch may substantially degrade the system performance, see~\cite{shokri2015mmWavecellular} and references therein.} which is employed in IEEE~802.15.3c and IEEE~802.11ad. The delay and coverage performance of these control channels are evaluated for a cellular context in~\cite{shokri2015mmWavecellular}, and evaluating their performance in short range networks is an interesting subject of future studies.

\section{Standardization in mmWave Communications}\label{sec: standardization}
\begin{figure}[!t]
\centering
  \includegraphics[width=0.8\columnwidth]{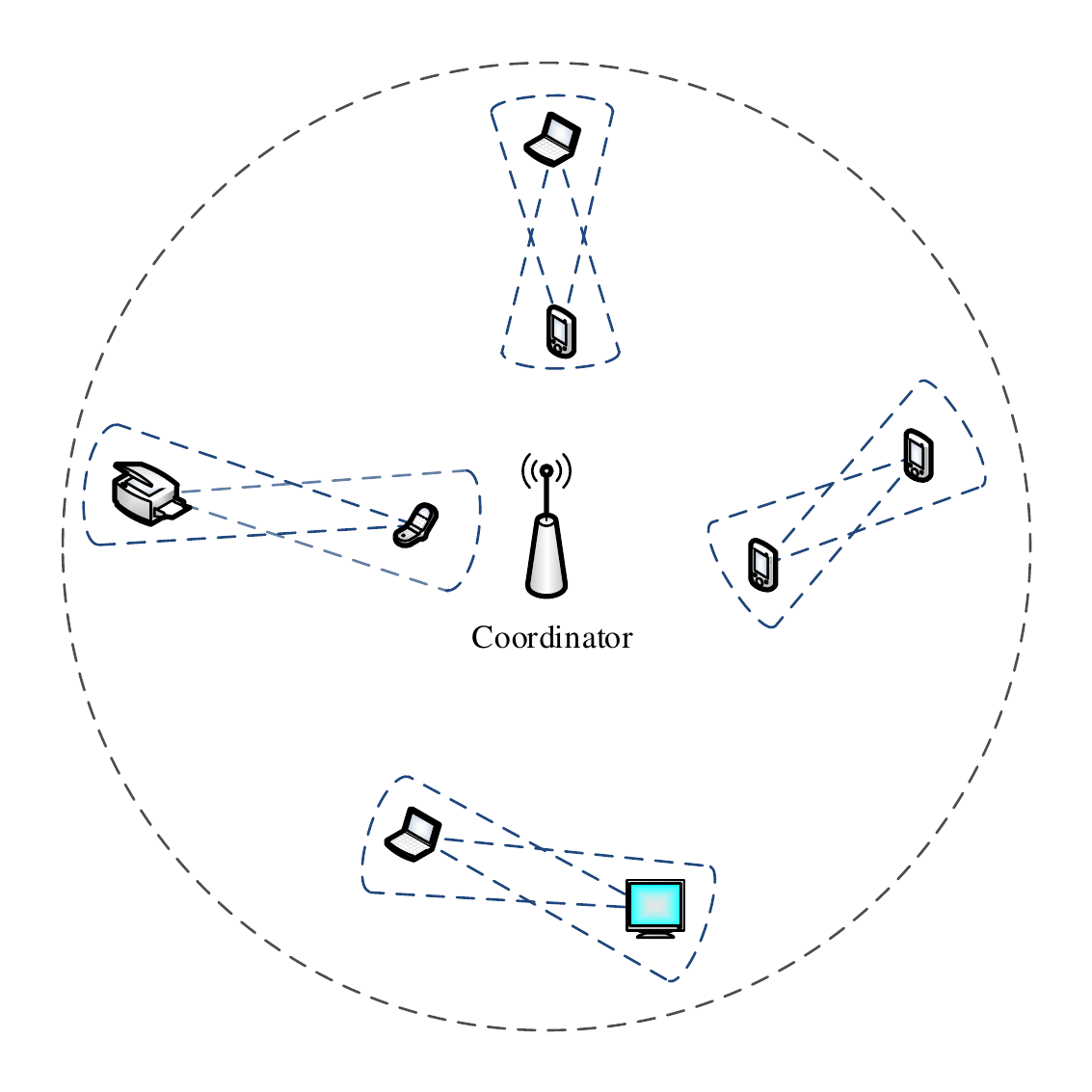}\\

  \caption{Network architecture of existing mmWave WPAN and WLAN. The coordinator broadcasts synchronization commands and manages available resources.}
  \label{fig: NetworkArchitecture}
\end{figure}

In this section, we shortly review the recent IEEE standards for personal and local area networks at 60~GHz.
Broadly speaking, the standards define a network with one coordinator and several mmWave devices.\footnote{ECMA~387 supports distributed network architectures as well~\cite{rappaport2014mmWaveBook}.} The coordinator, which can be a device itself, is responsible for broadcasting synchronization beacons and managing radio resources. Fig.~\ref{fig: NetworkArchitecture} shows a mmWave network with four directional links.

\subsection{Personal Area Networks: IEEE~802.15.3.c}
The IEEE~802.15.3c standard~\cite{rappaport2014mmWaveBook} has been considered as one of the prominent MAC candidates to support mmWave wireless personal area networks, known as piconets. Supporting up to 5.78~Gbps data rate,
it enables several applications such as high speed Internet access, streaming content, video on demand, and high definition TV.

Among a group of devices, one will be selected as piconet coordinator (PNC), broadcasting beacon messages. Time is divided into successive super-frames, each consisting of three portions: beacon, contention access period (CAP), and channel time allocation period (CTAP), as shown in Fig.~\ref{subfig: TimingStructure-802.15.3c}.
\NEW{In the beacon, the coordinator transmits an omnidirectional or multiple quasi-omnidirectional beacons to facilitate the discovery procedure.
In the CAP, devices contend to register their channel access requests at the PNC, based on carrier sense multiple access with collision avoidance (CSMA/CA). Although some devices with low QoS requirements may use this period for data transmission, PNC serves requests with high QoS demands, registered in CAP, during CTAP. Resource allocation in CTAP is based on time division multiple access (TDMA). CTAP is comprised of channel time allocations (CTAs), serving data traffic that requires QoS guarantees.}

\subsection{Local Area Networks: IEEE~802.11ad}
IEEE~802.11ad adds modifications to the IEEE~802.11 physical and MAC layers to enable mmWave communications at 60~GHz. It provides up to 6.7~Gbps data rate using 2.16~GHz bandwidth over a short range. IEEE~802.11ad supports many applications, including uncompressed high-definition multimedia transmissions and wireless docking stations.

\begin{figure}[!t]
	\centering
	\subfigure[Superframe of IEEE~802.15.3c]{
		\includegraphics[width=\columnwidth]{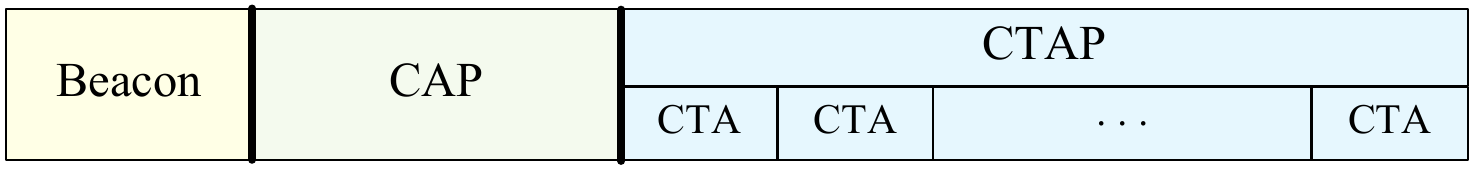}
		\label{subfig: TimingStructure-802.15.3c}
	}
	\subfigure[Beacon interval of IEEE~802.11ad]{
	\centering
		\includegraphics[width=\columnwidth]{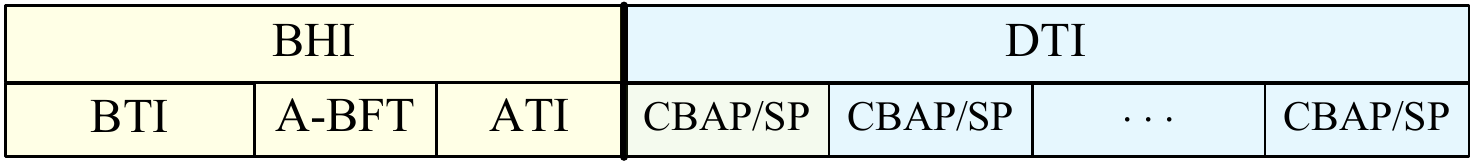}
		\label{subfig: TimingStructure-802.11ad}
	}		
	
    \caption{Network timing structure of existing IEEE mmWave standards. In IEEE~802.15.3c, beacon messages are transmitted in BP. Channel access requests are made in CAP and served in CTAP using TDMA. Similar procedures are adopted in IEEE~802.11ad.}
	\label{fig: TimingStructure}
\end{figure}

IEEE~802.11ad defines a network as a personal basic service set (PBSS) with one coordinator, called PBSS control point (PCP), and several stations.
\NEW{A superframe, called beacon interval, is divided into a beacon header interval (BHI) and a data transfer interval (DTI). BHI consists of a beacon transmission interval (BTI), an association beamforming training (A-BFT), and an announcement transmission interval (ATI). DTI consists of several contention-based access periods (CBAPs) and service periods (SPs). In BTI, PCP transmits directional beacon frames that contain basic timing for the personal BSS, followed by beamforming training and association to PCP in the A-BFT period. ATI is allocated for request-response services where PCP sends information to the stations.
Depending on the required QoS level, a device will be scheduled in the CBAP to transmit data using CSMA/CA, or in the SP for contention-free access using TDMA. This schedule is announced to the participating stations prior to the start of DTI.}
Fig.~\ref{subfig: TimingStructure-802.11ad} illustrates generic timing segmentation of a superframe in IEEE~802.15.3c and a beacon interval in IEEE~802.11ad.

\subsection{\NEW{Local Area Networks: IEEE~802.11ay}}
\NEW{IEEE 802.11ay is the most recent study group within IEEE, formed in May~2015, that aims to modify IEEE~802.11ad to enhance the throughput, range, and most importantly the use cases, while ensuring backward compatibility and coexistence with legacy mmWave standards. Supporting data rates of at least 20~Gbps\footnote{To the best of our knowledge, the maximal target data rate is not specified so far, but there are indications that the target rates aim towards hundred(s) of~Gbps} and a maximum range of 1000~m, IEEE~802.11ay enables a wide variety of applications ranging from backup wireless connections in data centers to mobile backhauling. To achieve these goals, the study group is investigating several techniques, including channel bonding, hybrid beamforming, and higher modulation orders, among others.
As the study group has not released any stable document so far, we cannot provide further details on this standard. In the next two sections, we highlight the bottlenecks of current mmWave standards and our suggestions to this study group on how to improve MAC layer efficiency of future mmWave networks.}

\section{\NEW{Limitations of Existing mmWave Standards}}\label{sec: Gap-Analysis}
In this section, we discuss the main MAC design issues that arise in mmWave communications and state the weaknesses of the current solutions, including existing standards, when they are applied to support next generation short range wireless communications.

\NEW{To highlight the existing challenges and possible solution approaches in the following sections, we simulated a mmWave WPAN with a random number of aligned mmWave links (aligned transmitter-receiver pairs),\footnote{\NEW{In many use cases of mmWave networks such as mobile backhauling, we can neglect the beam training overhead due to low-mobility and high traffic. The impact of the alignment overhead on the network performance is discussed in Section~\ref{sec: alignment-overhead}.}} all operating with the same beamwidth.
The number of links is a Poisson random variable with a given density per unit area. They are uniformly distributed in a 10x10~${\text{m}}^2$ area and operate at 60~GHz. We also uniformly distribute a random number of obstacles with density 0.25 (on average 1 obstacle in a 2x2~${\text{m}}^2$ area) in the environment. The obstacles are in the shape of lines with random orientation, and their length is uniformly distributed between 0 and 1~m. Penetration loss is -30~dB, path-loss exponent is 3, and the minimum required SNR at the receiver is 10~dB. We simulate both slotted~ALOHA and TDMA, a simple collision-based versus a simple collision-free protocol. Both slotted~ALOHA and TDMA use the same directionality level. For slotted~ALOHA, in a given time slot, every link will be active with a given transmission probability. For TDMA, we activate only one link at a time, similar to existing mmWave standards. Active links transmit with power 2.5~mW. Every transmitter generates traffic with constant bit rate (CBR) 300~Mbps, the size of all packets is 10~kB, time slot duration is 25~$\mu$s, transmission rate is 1 packet per slot (link capacity around 3~Gbps), the transmitters have infinite buffers to save and transmit the packets, and the emulation time is 1 second. For benchmarking purposes, we also simulate a network with omnidirectional communications, where we fix all the parameters and only increase the transmit power to achieve the same transmission range as directional communications.}

\subsection{\NEW{Transitional Behavior of Interference}}\label{sec: transitional-behavior}
Directional communications with pencil-beam operation significantly reduces multiuser interference in mmWave networks. An interesting question is whether in this case, a mmWave network is noise-limited, as opposed to conventional interference-limited networks. This is a fundamental question at the MAC layer that affects the design principles of almost all MAC layer functions. For instance, as the system moves to the noise-limited regime, the required complexity for proper resource allocation and interference avoidance functions at the MAC layer is substantially reduced~\cite{Shokri2015Transitional}.
Instead, pencil-beam operation complicates negotiation among different devices in a network, as control message exchange may require a time consuming beam training procedure between transmitter and receiver~\cite{shokri2015mmWavecellular}.
The seminal work in~\cite{Singh2011Interference} confirms the feasibility of a \emph{pseudowired} (noise-limited) abstraction in outdoor mmWave mesh networks. However, as shown in~\cite{Shokri2015Beam}, activating all links may cause a significant performance drop compared to the optimal resource allocation in dense deployment scenarios due to non-negligible multiuser interference. Further, the comprehensive analysis of~\cite{Shokri2015Transitional} illustrates that mmWave networks may not be necessarily noise-limited; rather they show a \emph{transitional behavior}, from a noise-limited to an interference-limited regime.

\begin{figure}[!t]
  \centering
  \subfigure[]{
    \begin{minipage}[c]{0.45\textwidth}
    \includegraphics[width=\columnwidth]{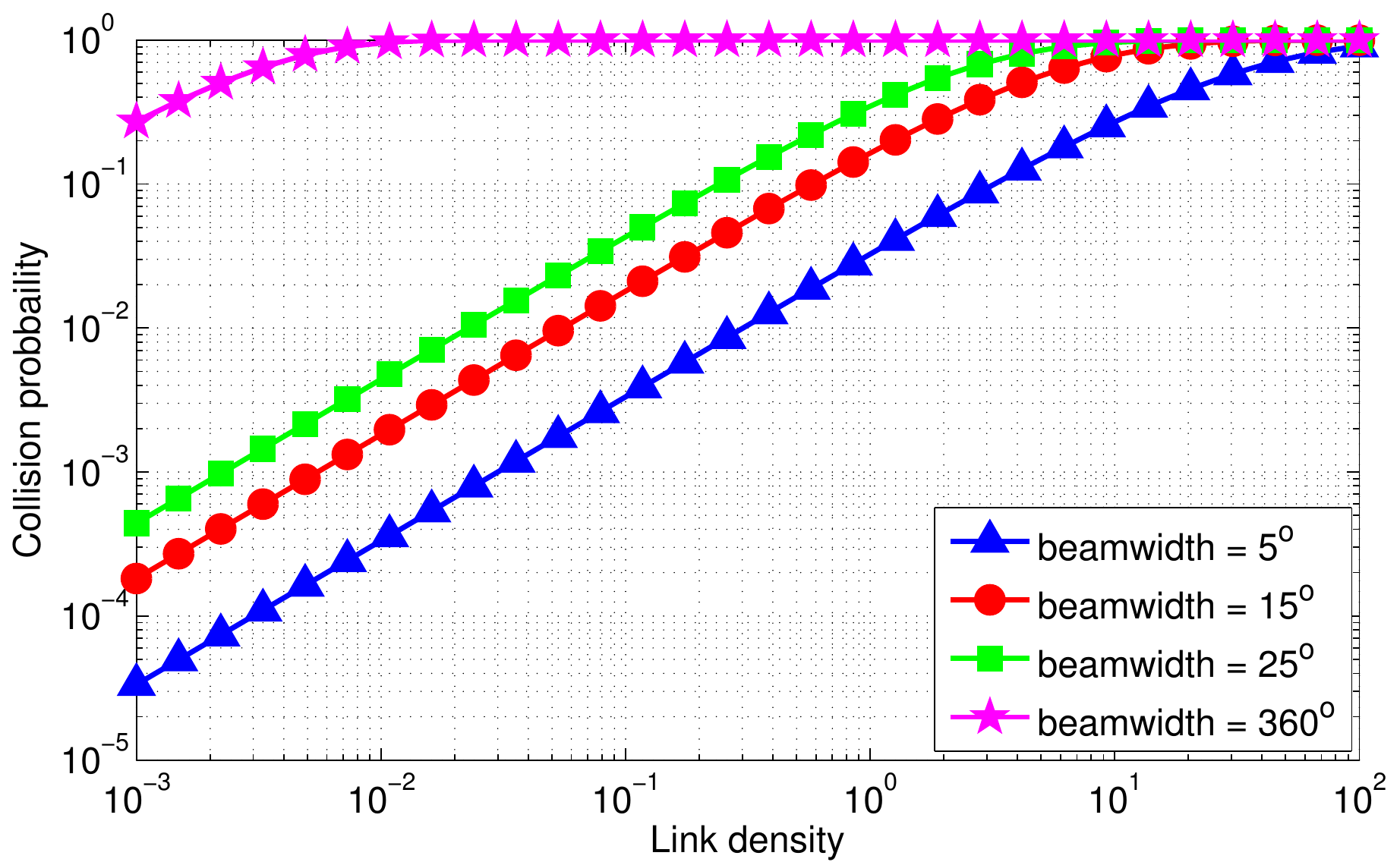}
	\label{subfig: CollisionGivenL_I_Theta}
    \end{minipage}%
  }
  \subfigure[]{
    \begin{minipage}[c]{0.45\textwidth}
    \includegraphics[width=\columnwidth]{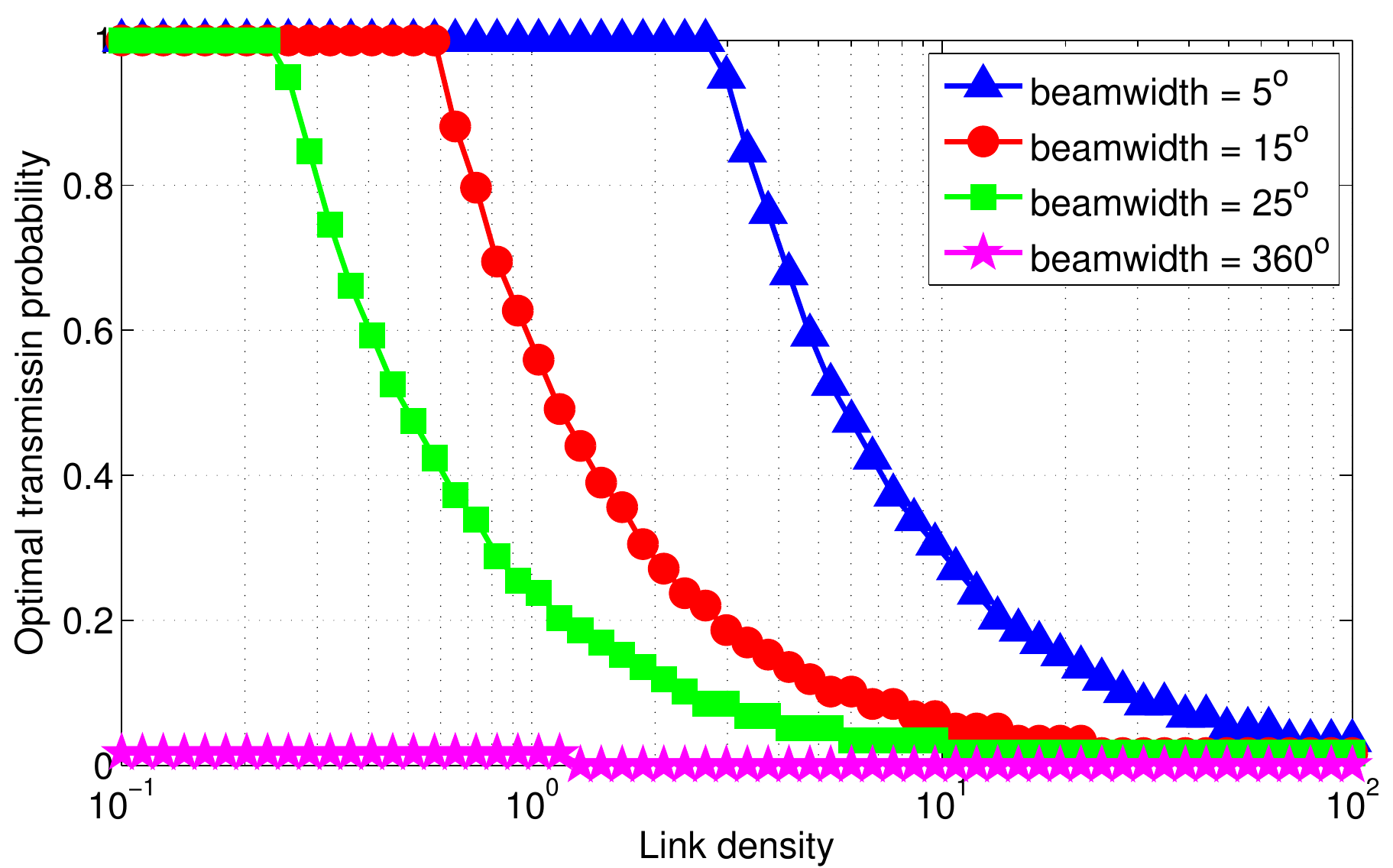}
	\label{subfig: OptimMACThr_OptimlTxProb}
    \end{minipage}%
  }

   \caption{Illustration of the transitional behavior of mmWave networks: \subref{subfig: CollisionGivenL_I_Theta} collision probability and~\subref{subfig: OptimMACThr_OptimlTxProb} optimal transmission probability of slotted ALOHA. \NEW{The negligible collision probability in~\subref{subfig: CollisionGivenL_I_Theta} and the very high optimal transmission probability in~\subref{subfig: OptimMACThr_OptimlTxProb} correspond to negligible multiuser interference. High collision probability and small optimal transmission probability correspond to the interference-limited regime. MmWave networks with narrow beam operation exhibit a full range of behaviors, from noise-limited to interference-limited, whereas microwave networks with omnidirectional operation always experience an interference-limited regime.}}
	\label{fig: Transitional-behavior}
\end{figure}
\NEW{Fig.~\ref{fig: Transitional-behavior} illustrates the transitional behavior of interference in a mmWave network.
Negligible collision probability in this figure indicates negligible multiuser interference, whereas high collision probability corresponds to the interference-limited regime. From Fig.~\ref{subfig: CollisionGivenL_I_Theta}, we see that even for a network of modest size, the collision probability may be high enough to invalidate the assumption of being in a noise-limited regime, e.g., 0.2 collision probability for the case of 1 transmitter in a 2x2~${\text{m}}^2$ area and an operating beamwidth of 25$\degree$. Moreover, as can be observed in all curves of Fig.~\ref{subfig: CollisionGivenL_I_Theta}, there is a transition from a noise-limited to an interference-limited regime in a mmWave network with directional communications, whereas traditional networks with omnidirectional communications always experience an interference-limited regime without any transitional behavior under ``realistic'' parameter choices.
The transitional region of mmWave networks depends on the density of the transmitters, the density and the average size of the obstacles, the operating beamwidth, and also the MAC protocol.}

Fig.~\ref{subfig: OptimMACThr_OptimlTxProb} shows the behavior of the optimal transmission probability that maximizes the throughput of slotted ALOHA as a function of link density and operating beamwidth. From the figure, it can be observed that the optimal transmission probability is 1 in many cases, implying that we can simply activate all links with no penalty for the average link throughput (noise-limited regime). However, as the operating beamwidth or the link density increases, we should activate fewer links by reducing the transmission probability, in order to decrease the high contention level inside the network (interference-limited regime).

\subsection{\NEW{Control and Data Rate Mismatches}}\label{sec: rate-mismatch}
Current collision avoidance mechanisms suggest that a network with uncoordinated users will benefit from accepting collisions on tiny signaling messages such as request-to-send (RTS) and clear-to-send (CTS) to avoid retransmission of large data messages. To increase the robustness of signaling messages, current mmWave standards transmit control messages at much lower rate compared to the data messages. IEEE~802.11ad, for instance, supports a peak transmission rate of 27.7~Mbps for control and 6.7~Gbps for data messages~\cite{Nitsche2014IEEE}.
\NEW{This significant mismatch between the transmission rates of control and data messages substantially increases the cost of collision avoidance procedures and challenges the efficacy of current mmWave standards in handling short packets. To illustrate this inefficiency, we provide the following example.}

Let $t_i$ be the time required to transmit message $i$. With negligible propagation and queuing delays and with no collision on a directed spatial channel, the current CSMA/CA protocol introduces the following delay to transmit a payload: $2 t_{\rm SIFS} + t_{\rm RTS} + t_{\rm CTS} + t_{\rm DIFS} + t_{\rm DATA}$, where $t_{\rm DATA} = t_{\rm header} + t_{\rm payload}$. Note that the transmitter should wait for a SIFS duration before sending every RTS and CTS, and wait for a DIFS duration before every regular data frame. In IEEE~802.11ad, $t_{\rm SIFS} = 2.5$~$\mu$s and $t_{\rm DIFS} = 6.5$~$\mu$s. Considering 20 Bytes for RTS and CTS messages, we have $t_{\rm RTS}=t_{\rm CTS}= 5.5$~$\mu$s. Every data packet contains an 8-Byte header, which should be transmitted at rate 27.7~Mbps, so $t_{\rm header}= 2.2$~$\mu$s. To transmit 10 KBytes of payload, we need only $t_{\rm DATA} = 13.6$~$\mu$s, while the total delay is 36.1~$\mu$s, leading to 37\% channel utilization. This inefficiency increases even more as the size of the payload reduces, for instance, the channel utilization is only 12\% for 1 KByte of payload. This means that CSMA/CA consumes around 90\% of the time resources only to ensure avoidance of collisions even in a noise-limited scenario. This inefficient handling of short packets hinders the applicability of current mmWave technologies (with Gbps data rate and small interference footprint) to massive wireless access scenarios where we have frequent transmissions of packets with small payloads. In fact, the huge overhead of having an unnecessary proactive collision avoidance protocol may be one of the main bottlenecks of future applications of mmWave networks.

The significant mismatch between transmission rates of control and data messages, along with the reduced average collision probability in mmWave networks, demands development of new MAC layer protocols with on-demand and minimal use of signaling. Note that proactive transmission of some vital control messages, such as beam training pilots, may still be mandatory. These mandatory control overheads may limit the delay/channel utilization performance and therefore the applicability of mmWave networks to use cases with sporadic transmissions of small payloads. \NEW{This suggests the existence of a minimal payload size to make the establishment of a costly mmWave link beneficial, whose characterization is an interesting topic for future studies.}

\subsection{Prolonged Backoff Time}\label{sec: prolonged-backoff-time}
Suppressing interference in mmWave networks with pencil-beam operation comes at the expense of complicated link establishment. Besides the huge overhead due to the collision avoidance procedures, conventional CSMA/CA that was originally developed for omnidirectional transmissions introduces a \emph{prolonged backoff time} in mmWave networks. To elaborate, assume that a mmWave transmitter tries to access the channel by sending an RTS message after the backoff timer expires. Assume that the receiver does not hear the RTS due to either deafness or blockage, and therefore does not send the CTS message. The traditional CSMA/CA protocol assumes that a collision occurred and therefore increases the backoff time exponentially. In mmWave networks, this may be the wrong decision, which may unnecessarily prolong the backoff time. Similar issues may also exist in the random access phase of mmWave cellular networks, as mentioned in~\cite{shokri2015mmWavecellular}.

To enhance the performance of CSMA/CA in directional communications, \cite{Ramanathan2005Adhoc} modifies traditional CSMA/CA such that each device exponentially increases the contention window size upon a missing ACK, while this increment is linear with each missing CTS. Although this proposal is better than the original CSMA/CA in the sense that different events demand different actions, it fails to solve the prolonged backoff time problem in mmWave systems. In fact, blockage, deafness, and collision, which are caused by different physical reasons, deserve a different handling at the MAC layer, a fact that is somewhat ignored in~\cite{Ramanathan2005Adhoc}. In the next section, we propose a novel MAC level message to facilitate the detection of a collision, thereby solving the prolonged backoff time problem.

\subsection{Alignment Overhead}\label{sec: alignment-overhead}
The adopted beam training approach of the existing standards introduces an alignment overhead, which depends on the number of directions that have to be searched, which in turn depends on the selected transmission and reception beamwidths. For a given beamwidth,~\cite{li2013efficient} suggests a new technique based on Rosenbrock search as a replacement for the existing two-stage exhaustive search, to reduce the alignment overhead by up to 65\% for a given operating beamwidth.

Alignment overhead, besides demanding more efficient search procedures, introduces an alignment-throughput tradeoff that necessitates an optimization over the operating beamwidth~\cite{Shokri2015Beam}. Narrower beamwidths increase the search granularity, thus the alignment overhead, but provide a higher transmission rate due to higher antenna gains and lower multiuser interference. Adopting larger beamwidths speeds up the search process at the expense of a degraded transmission rate. The tradeoff shows that using extremely narrow beams (or excessively increasing the beamforming codebook size) is not beneficial in general due to the increased alignment overhead, and there is an optimal beamwidth (optimal codebook size) at which the tradeoff is optimized~\cite{Shokri2015Beam}.

\section{\NEW{MAC Design for Future Short Range mmWave Networks}}\label{sec: MAC-Design-Aspects}
\NEW{In this section, we discuss the implications of the fundamental aspects highlighted in the previous section on the efficient MAC design in future mmWave networks.}

\subsection{\NEW{Collision-aware Hybrid MAC}}\label{sec: hyybrid-MAC}
\begin{figure}[!t]
  \centering
  \subfigure[]{
    \begin{minipage}[c]{0.45\textwidth}
    \includegraphics[width=\columnwidth]{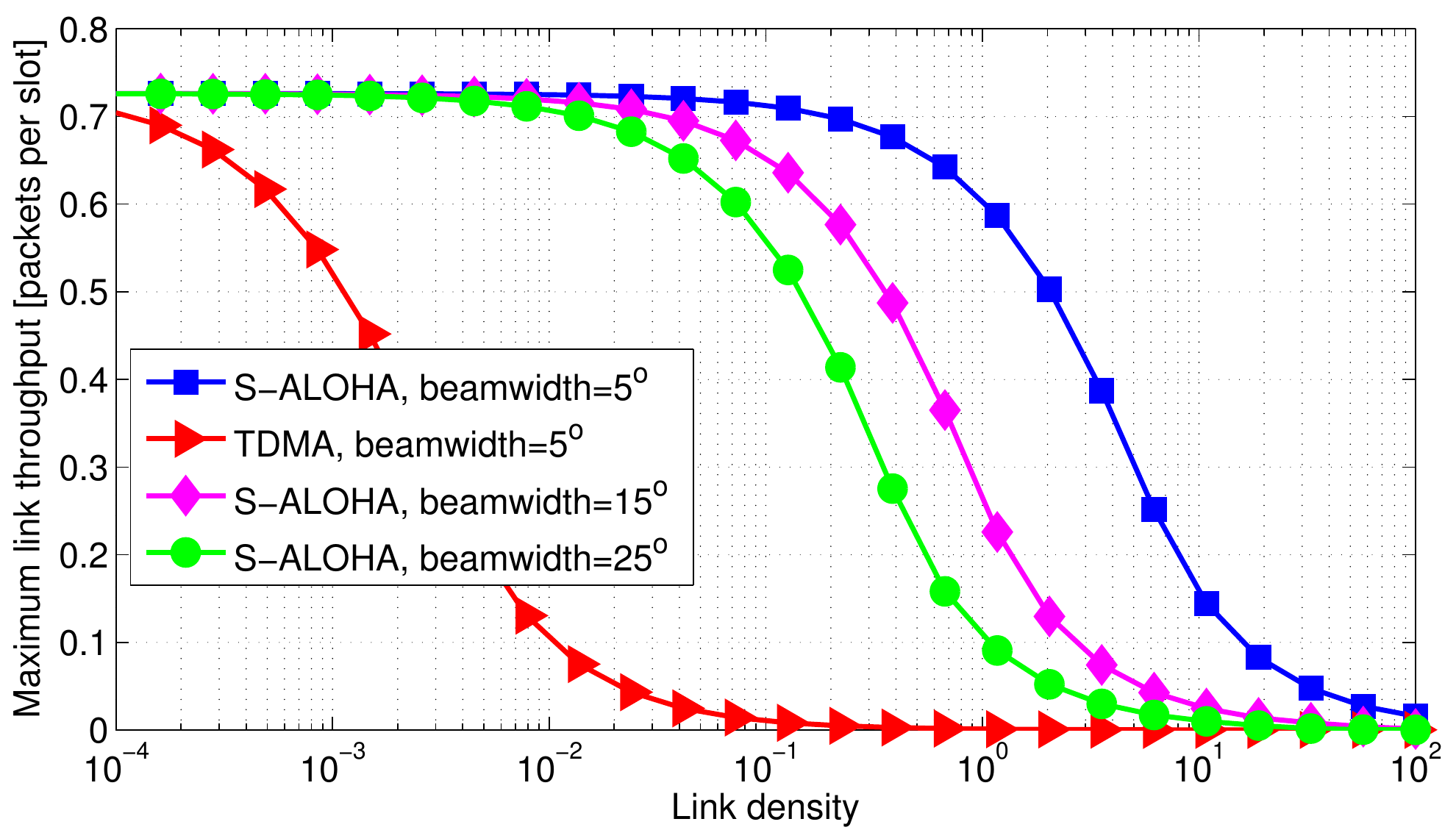}
	\label{subfig: Comparison3}
    \end{minipage}%
  }
  \subfigure[]{
    \begin{minipage}[c]{0.45\textwidth}
    \includegraphics[width=\columnwidth]{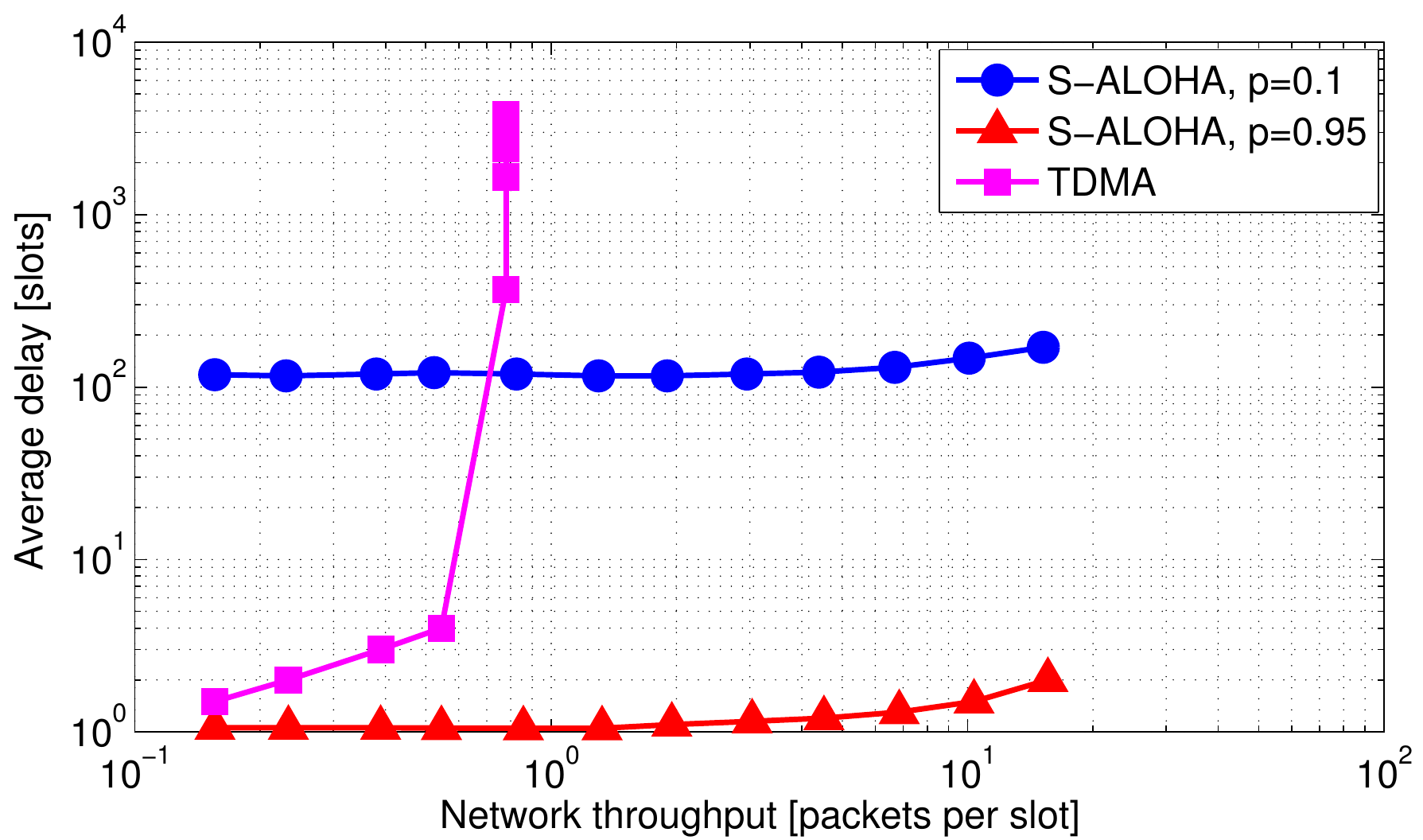}
	\label{subfig: Comparison4}
    \end{minipage}%
  }

   \caption{Performance comparison of slotted~ALOHA and TDMA in mmWave WPANs. The alignment overhead is neglected. ``S-ALOHA" stands for slotted~ALOHA, and $p$ is its transmission probability. Different points of \subref{subfig: Comparison4} represent different link densities from 0.02 to 2 links per unit area. Operating beamwidth in \subref{subfig: Comparison4} is 10$^{\rm o}$. Increasing the link density may reduce the link throughput, increase the network throughput, and increase the delay. Slotted~ALOHA significantly outperforms TDMA in terms of link throughput, network throughput, and delay performance. On the other hand, TDMA guarantees collision-free communication.}
	\label{fig: SlottedALOHA-vs-TDMA}
\end{figure}
\NEW{To investigate proper resource allocation strategies for mmWave networks, we compare the average throughput of a link, the network throughput, and the delay performance of slotted~ALOHA to those of TDMA in Fig.~\ref{fig: SlottedALOHA-vs-TDMA}. We define delay as the difference between the time a new packet is inserted into the transmission queue at the transmitter and the time it is correctly received at the receiver.}
Specifically, Fig.~\ref{subfig: Comparison3} reports the maximum throughput of a link in slotted~ALOHA, associated with the optimal transmission probability in Fig.~\ref{subfig: OptimMACThr_OptimlTxProb}, and Fig.~\ref{subfig: Comparison4} shows the network throughput against the corresponding average delay obtained by changing the link density. First, neglecting the alignment overhead, the throughput of a link in slotted~ALOHA will decrease with the operating beamwidth, due to a higher collision probability.
Moreover, TDMA activates only one link at a time -- orthogonal use of time resources -- irrespective of the number of links. Considering the traffic generation rate of this example, which is 0.1 of the link capacity, the network will be saturated roughly with 10 links, and further increasing the number of links will not improve the network throughput (see Fig.~\ref{subfig: Comparison4}),\footnote{The network throughput of TDMA is at most 1 packet per slot. This upper bound is achieved if there is no obstacle in the environment.} but will instead reduce the time share of every link and consequently reduce the average throughput of a link, see Fig.~\ref{subfig: Comparison3}. Besides, every link experiences a higher delay to access the channel and transmit its data, see different points of the TDMA curve in Fig.~\ref{subfig: Comparison4}. Note that with a fixed packet generation rate, the \emph{effective link capacity} (link capacity multiplied by its time share) of every link in TDMA decreases with the number of links in the network, so the queues of the transmitter may become unstable. \NEW{Instead, slotted~ALOHA leverages small multiuser interference and is able to effectively re-use the time resources (spatial gain), thus every link can handle more traffic due to a higher effective link capacity. Significant spatial gain in mmWave networks is also highlighted in~\cite{son2012frame,niu2015exploiting}, where the authors try to leverage this gain in a general noise-limited network~\cite{son2012frame} and in a device to device network~\cite{niu2015exploiting}.
Fig.~\ref{subfig: Comparison4} shows that slotted~ALOHA significantly outperforms TDMA in terms of both network throughput and delay, thanks to this significant spatial gain. However, unlike slotted~ALOHA, TDMA can guarantee communication with no collisions, which may be of importance in some applications, e.g., under short delay or ultra high reliability constraints.}

Current mmWave standards, such as IEEE~802.15.3c and IEEE~802.11ad, adopt resource allocation approaches that were originally developed for interference-limited microwave networks. In particular, the network traffic is mostly served in the contention-free phase even in a noise-limited regime. \NEW{However, devices in a mmWave network may show a full range of behaviors from noise-limited to interference-limited, demanding a dynamic (collision-aware) incorporation of both contention-based and contention-free phases in the resource allocation framework. The contention-based phase improves the throughput/delay performance by leveraging concurrent transmissions, and the contention-free phase can be applied to deliver only the remaining traffic. To exemplify the superior feature of this collision-aware hybrid MAC, we note that isolated devices that receive almost no interference can transmit during all the data transmission interval (DTI) without extra scheduling delay, whereas existing hybrid MAC solutions force them to register their requests, wait until they are scheduled, and transmit for a short portion of DTI, see Fig.~\ref{fig: TimingStructure}.
It follows that, in a noise-limited regime, we may deliver most of the traffic in the contention-based phase (where contention does not actually occur). This can lead to providing around an order of magnitude higher throughput for a given link density and supporting an order of magnitude denser network with given average per-link throughput, see Fig.~\ref{subfig: Comparison3}, extending the use cases of future mmWave networks.}

\subsection{\NEW{Efficient Retransmission Policy}}\label{sec: retransmission-policy}
\NEW{In the previous subsection, we showed that the TDMA phase of the hybrid MAC of existing standards needs modification. In this subsection, we show that the CSMA/CA phase of their hybrid MAC needs to be thoroughly modified as well.}

\NEW{Retransmission after a random backoff is a common solution to handle collisions without any network-wide coordination.
In CSMA/CA, adopted by existing standards, retransmission of an RTS message after random backoff leads to a virtual channel reservation and collision-free data transmission. It can also alleviate the well-known hidden and exposed node problems. However, the special characteristics of mmWave networks diminish the benefits of CSMA/CA over simple CSMA. First, proactive channel reservation generally causes a significant throughput drop and extra delay in mmWave networks due to the overwhelming signaling overhead, as described in Section~\ref{sec: rate-mismatch}. Moreover, the directionality of mmWave networks substantially reduces the hidden and exposed node problems, and consequently the need for collision avoidance. In the following, we also show that different links experience different collision levels, a feature that should be addressed in designing proper retransmission policies.
}

\begin{figure}[!t]
	\centering
  \subfigure[operating beamwidth 5$\degree$]{
    \begin{minipage}[c]{0.48\textwidth}
    \includegraphics[width=\columnwidth]{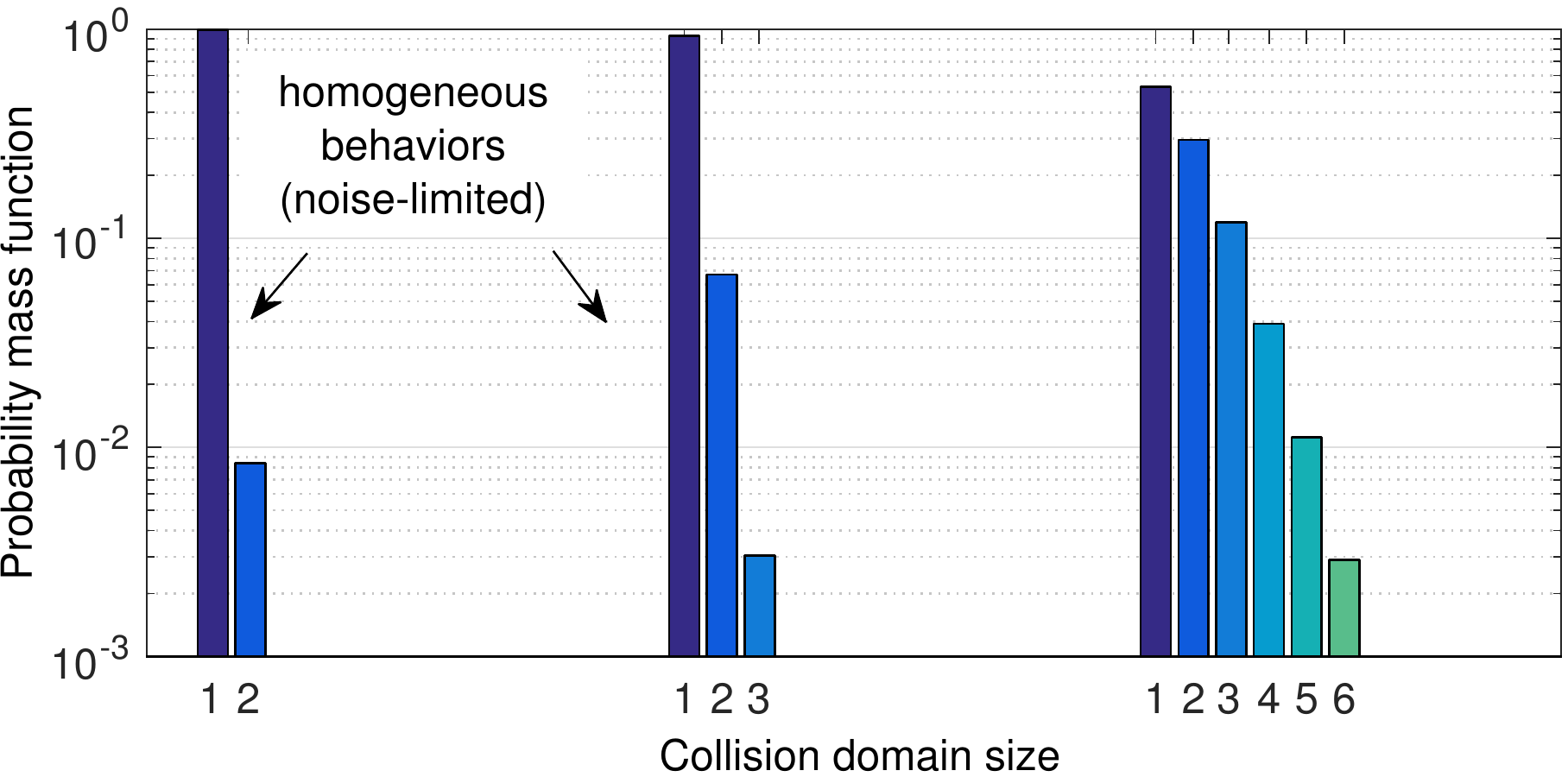}
    \label{subfig: DegreeDist_Beamwidth_5}
    \end{minipage}%
  }
\vspace{+5mm}
 \subfigure[operating beamwidth 30$\degree$]{
    \begin{minipage}[c]{0.48\textwidth}
    \includegraphics[width=\columnwidth]{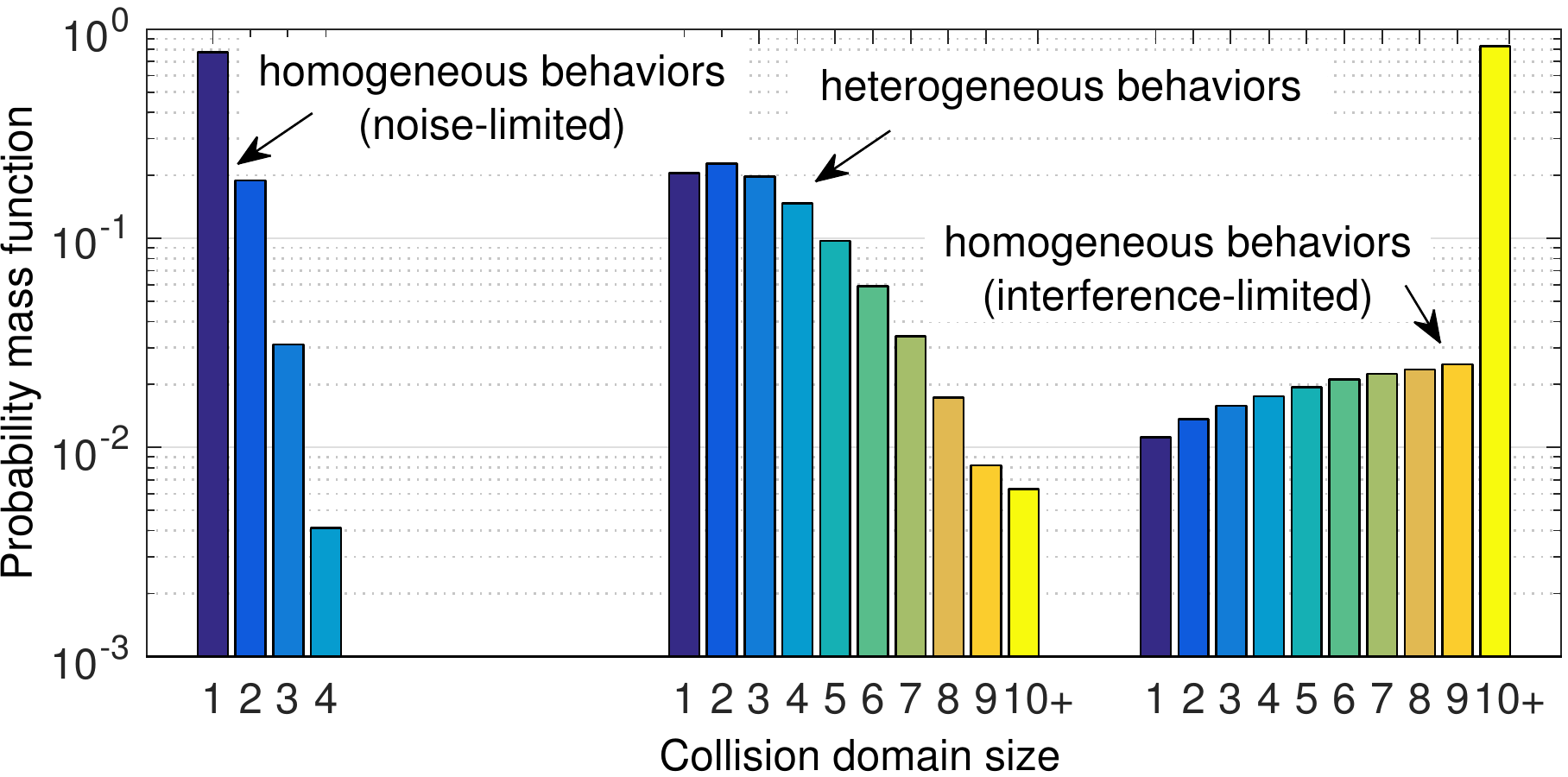}
    \label{subfig: DegreeDist_Beamwidth_30}
    \end{minipage}%
  }
\vspace{+5mm}
\subfigure[operating beamwidth 360$\degree$]{
    \begin{minipage}[c]{0.48\textwidth}
    \includegraphics[width=\columnwidth]{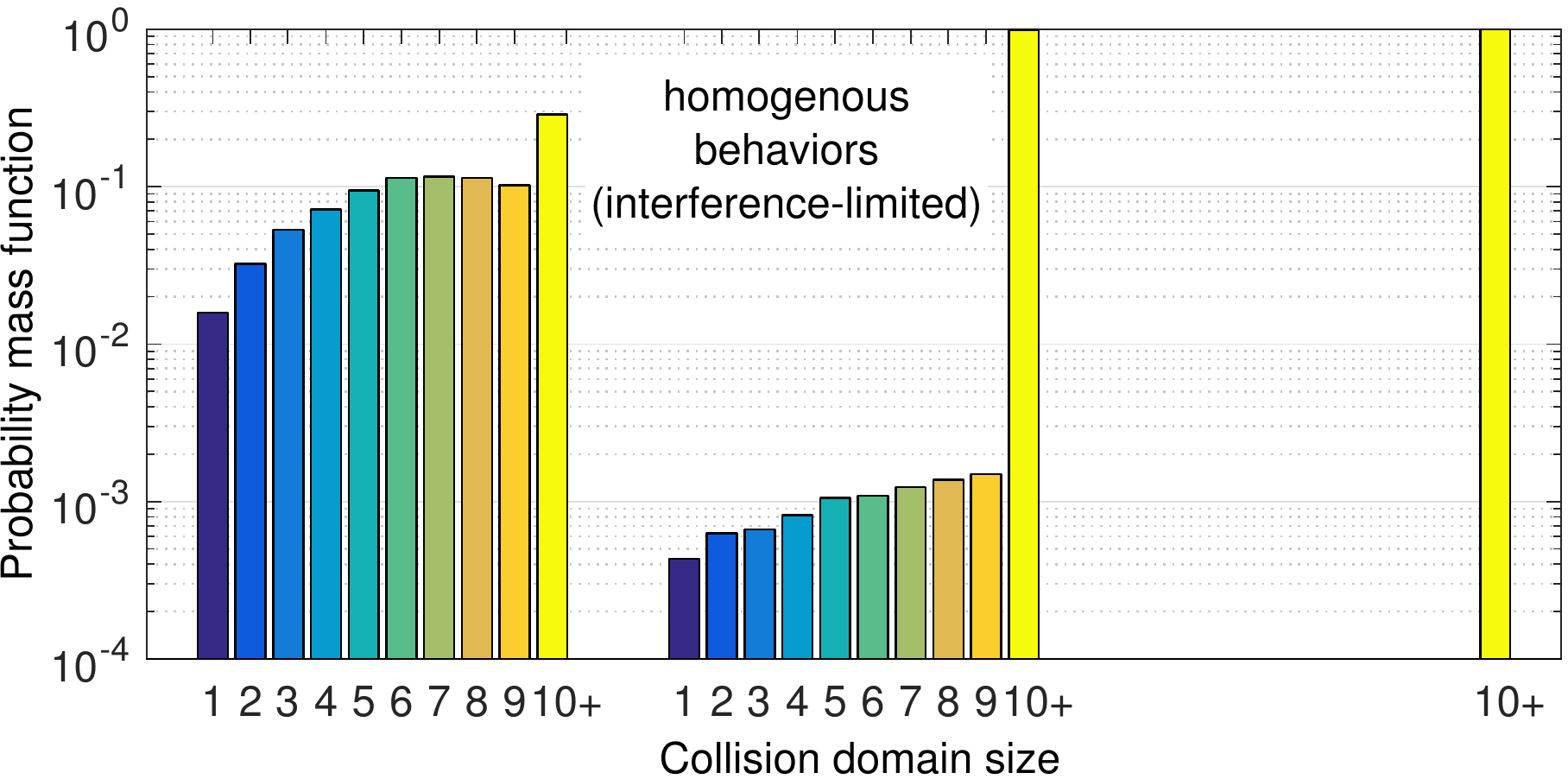}
    \label{subfig: DegreeDist_Beamwidth_360}
    \end{minipage}%
  }	

\caption{\NEW{Distribution of the number of conflicting links for different operating beamwidth and density of the transmitters. Three sets of distributions in each figure from left to right correspond to link densities of 0.11, 1, and 10 links per square meter, respectively. Obstacle density is 0.11. Size 1 for collision domain represents isolated devices with no incoming interference.}}
	\label{fig: CollProb_Densities}
\end{figure}

\NEW{The design of efficient retransmission policies depends largely on the distribution of the number of links in the same collision domain (links with strong mutual interference). An increased number of links in the same collision domain results in more retransmissions, and therefore a higher delay to establish a channel. Fig.~\ref{fig: CollProb_Densities} shows such distribution as a function of operating beamwidth and link density. Each plot contains three sets of distributions that correspond to link densities of 0.11, 1, and 10 links per square meter, from left to right, respectively. Under pencil-beam operation and relatively low link density, a mmWave network is comprised of devices with homogenous collision behavior, as almost all of them show a noise-limited behavior. Increasing either the link density or the operating beamwidth shifts the mmWave network toward the interference-limited regime. In the extreme case of omnidirectional communication, all the devices show another homogenous behavior, i.e., an interference-limited regime. However, as can be observed in Fig.~\ref{subfig: DegreeDist_Beamwidth_30}, devices in mmWave networks may show a full range of behaviors from noise-limited to interference-limited. To design an efficient retransmission policy for such networks, a link should be able to identify the size of the collision domain it belongs to. This is a largely open problem in mmWave networks, demanding new analytical models and protocol designs. A direct research question is whether, in mmWave networks, reactive retransmission of a data message after a random backoff procedure (CSMA) is a better option to be adopted by all devices than proactive execution of costly collision avoidance mechanisms (CSMA/CA). Another open question is whether, upon detecting a collision, the transmitter-receiver pairs should (1) adopt a narrower beamwidth at the cost of some extra alignment overhead but with the possible benefit of operating with no multiuser interference and therefore significant throughput enhancements (see Section~\ref{sec: hyybrid-MAC}), (2) execute a random backoff procedure to share DTI among the set  of colliding links in a distributed fashion, or (3) send a TDMA reservation request to the coordinator. The proper choice depends on the use case, QoS requirements, and available information such as the collision domain size.}

\subsection{\NEW{Collision Notification}}\label{sec: hybrid-control-plane}
\NEW{Due to the heterogenous behavior of the collisions in mmWave networks, detecting the collision level provides useful information for a link to adopt proper retransmission policies, make control plane more efficient, implement an on-demand TDMA phase, and solve the prolonged backoff time problem.}

\NEW{To develop a procedure that estimates the collision level, we first consider orthogonal signatures for different types of messages like RTS, CTS, and data. Inspired by the use of pseudo-orthogonal symbol sequences (PSS) in synchronization symbols, we can readily implement orthogonal signatures by adding corresponding PSSs to the header of any message. Then, a correlator at any receiver matches (the time shifts of) the received signal with the reference symbol sequences to identify the type of the received messages. First, as this scheme is very robust and can work well even at very low SNRs, we can transmit this part of the header at very high rate, decreasing the time overhead of this part of the header. Moreover, if multiple messages of the same type are received (due to multiple transmitters), the receiver can distinguish them as they are received by different time shifts. If messages of different types are received, again, the receiver can distinguish them due to their orthogonal signatures. Note that the types of the superimposed messages are detectable due to the robustness of PSSs; they are short and easily detectable even at very low SNR, at which neither the header nor the payload are decodable.}

\NEW{We introduce a novel MAC level message, called collision notification (CN), which any receiver will transmit upon receiving messages that are not decodable due to a collision. To distinguish non-decodable message(s) due to a collision from those due to severe channel attenuations, we note that the correlator's output at the receiver peaks at several time shifts in the case of collision. Alternatively, the receiver can use a simple hard decision based on the received energy (energy detector); the level of the received power is very low in the case of severe channel attenuation, blockage, or deafness, whereas it is very high in the case of collision with multiple simultaneous received signals. The proposed CN message can address the prolonged backoff time problem, facilitate the on-demand realization of the TDMA phase, and reduce the frequency of unnecessary executions of the costly collision avoidance procedure. In the following, due to lack of space, we only mention how the CN message alleviates the prolonged backoff time problem, whereas other direct applications of the CN message are the subject of our future work.}

\begin{figure}[!t]
  \includegraphics[width = \columnwidth]{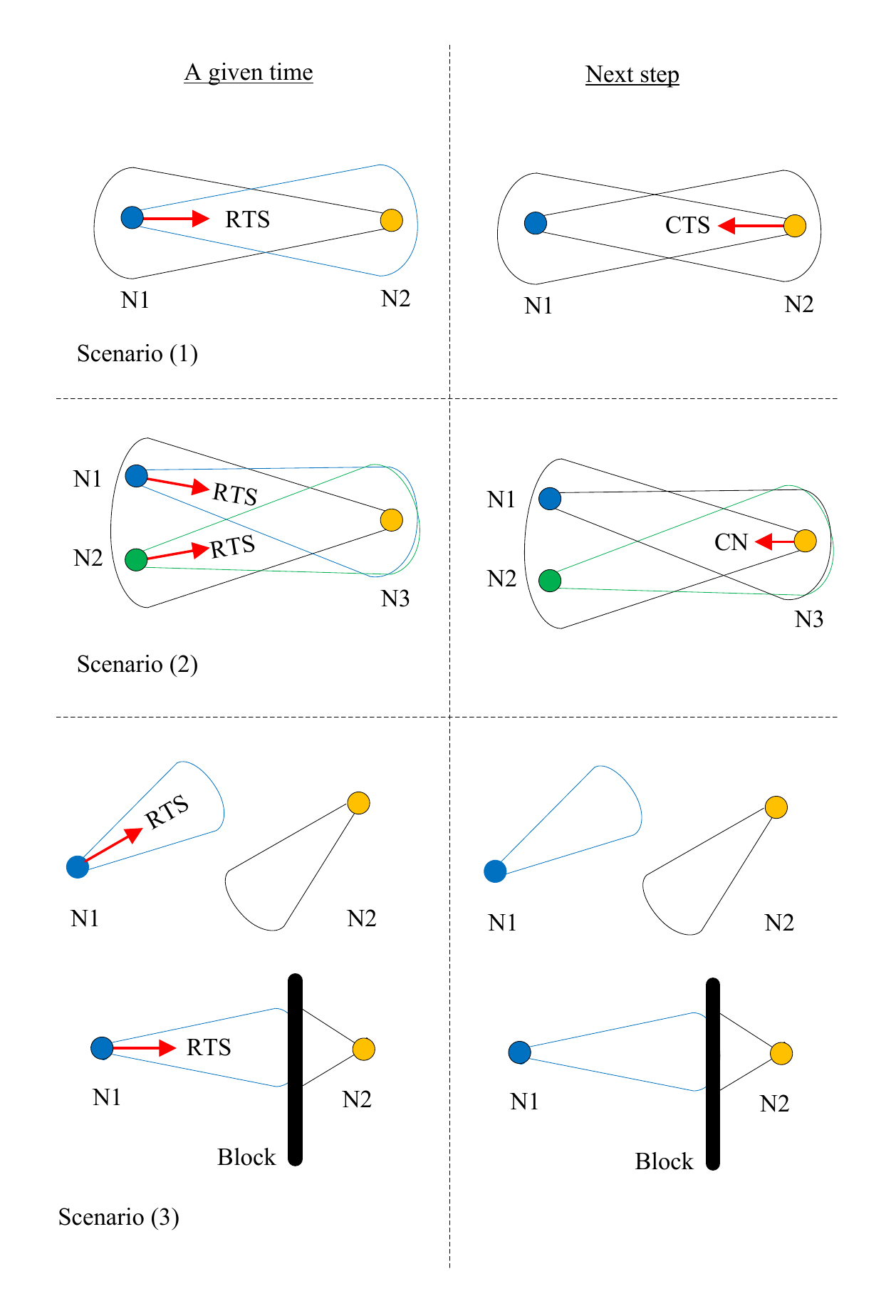}

  \caption{A simple protocol for mitigating prolonged backoff time.
  For a given time, in Scenario 1, device N2 detects an RTS. The next step is for device N2 to send a CTS message to reserve the channel. In Scenario 2, device N3 receives more than one RTS at the same time. It sends a CN message to let the transmitters run the backoff procedure. In Scenario 3, device N2 does not receive the RTS of device N1 due to either deafness or blockage, and will be silent at the next step.
  }
  \label{fig: Protocol}
\end{figure}

A simple scheme to alleviate the prolonged backoff time problem, illustrated in Fig.~\ref{fig: Protocol}, may work as follows. After sending a directional (or omnidirectional) RTS to a receiver that is ready to receive, the following cases might occur:
\begin{itemize}
  \item Scenario 1 (success): The transmitter receives a CTS before timeout. Then, it starts transmission based on the CSMA/CA mechanism.
  \item Scenario 2 (collision): The receiver fails to decode the RTS due to a collision. It sends a CN message. Upon receiving the CN message, the transmitter knows that there is a contention to access this channel in this direction, and therefore sends another RTS after running the random backoff procedure.
  \item Scenario 3 (deafness or blockage): The transmitter does not receive a CTS nor a CN. In this case, after timeout, it knows that there is either deafness or blockage. Hence, it tries to find another directed spatial channel instead of running an unnecessary backoff.
\end{itemize}
\begin{figure}[!t]
  \centering
  \includegraphics[width=\columnwidth]{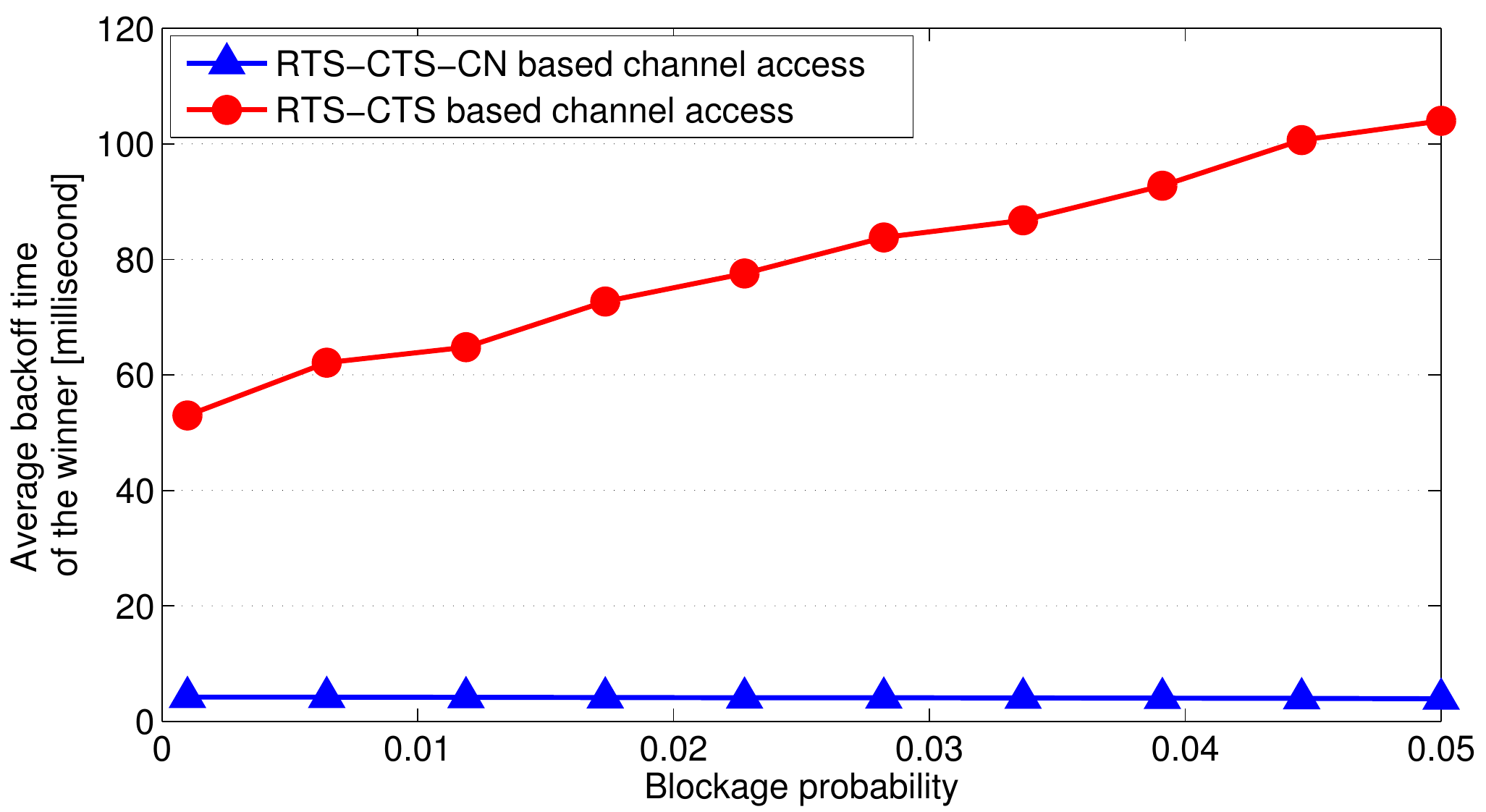}

  \caption{Average backoff time of the device winning the contention among 20 devices for accessing the same transmission resource (frequency and direction). Standard RTS-CTS based negotiation leads to unnecessarily prolonged backoff time, while a slight modification of this standard negotiation, by introducing CN, effectively mitigates the problem.
  }
  \label{fig: PerformanceComparision_Blockage}
\end{figure}
\NEW{If a set of receivers fail to decode RTS messages, they will all respond back with CN messages in the same direction they were listening, and their intended transmitters will then execute a collision avoidance procedure. If a set of transmitters correctly receive CN messages, they will start the conventional collision avoidance procedure (Scenario~2). If the CN messages collide, the intended transmitters can still identify the existence of multiple CN messages, even if their entire headers/payloads are not decodable, so the corresponding transmitters correctly execute the collision avoidance procedure. The CN message, however, increases feedback traffic.
The minimum required size of PSSs, required SNR, transmission rate of PSSs, and the performance of the message type detection will be the subject of our future studies.}

Thanks to the CN message, the transmitter can sense the presence of contention in the channel and take the proper MAC layer action to avoid the prolonged backoff time, which is the result of deafness and blockage, and not of contention on the channel. We simulate a network with a Bernoulli link failure model, i.e., every link fails due to blockage independently and with constant blockage probability. Fig.~\ref{fig: PerformanceComparision_Blockage} shows the performance enhancement due to the introduction of CN. With a blockage probability of 0.02, for instance, the average backoff time will be dramatically decreased by about 95\% (twenty times) if CN is used.

\subsection{\NEW{Multihop Communications}}
\NEW{Relaying and multihop communications are key components of future mmWave networks for range extension and for blockage alleviation~\cite{shokri2015mmWavecellular,rappaport2014mmWaveBook,kim2013joint,Singh2009Blockage}. It is also essential for multihop backhauling, which is an important use case of IEEE~802.11ay. In~\cite{kim2013joint}, range extension using a relay node is investigated for an outdoor sport broadcasting system. Extensive analysis demonstrated that high quality live videos of 10 sources can be efficiently transmitted over 300~m. Besides range extension,~\cite{Singh2009Blockage} showed that having an alternative path using relay node(s) can significantly alleviate blockage. The backup paths are recorded in the coordinator and established upon blockage on the direct path, increasing connectivity to about $100\%$.}

\NEW{Unfortunately, current mmWave standards support only single- or two-hop links\footnote{IEEE~802.15.3c supports only single-hop communications, while ECMA~387 and IEEE~802.11ad support also two-hop communications.} rather than the complete multihop communication capability envisioned in IEEE~802.11ay.
Adding more hops entails additional alignment overhead per hop, which may limit the benefits of multihop communications. As stated in~\cite{shokri2015mmWavecellular}, the beamforming vector of the analog beamformer depends only on the large scale components of the channel, which will be almost constant over many consecutive superframes (beacon intervals). However, current mmWave standards neglect this important feature and perform a complete beam training procedure in every superframe. We suggest that each device estimates the topology of the network in the neighbor discovery phase. Then, it creates a table of proper spatial resources (directions) based on the feedback received from previous transmission attempts (piggybacking over data transmissions). The table is updated upon every received feedback, and each transmitter tries to communicate with other devices using the most updated table. This a priori information on the possible directions can substantially reduce the beam training space, thereby reducing the alignment overhead. The design of the analog beamformer is then reduced to beam-tracking over consecutive superframes, while the digital beamformer (in a hybrid beamforming architecture) may be still designed per superframe.}

\NEW{In addition to more efficient beam training, a joint routing and scheduling approach is necessary in multihop communications to leverage the low interference footprint in mmWave communications using scheduling, while guaranteeing connectivity using routing protocols. Designing such joint approach is an interesting future research direction.}

\section{Conclusions}\label{sec: concluding-remarks}
Millimeter wave (mmWave) communication systems are promising solutions to provide extremely high data rates and support massive uncoordinated access in future wireless networks. Severe channel attenuation, blockage, and deafness, along with a reduced interference footprint, differentiate mmWave systems from legacy systems that operate at microwave frequencies. \NEW{MmWave networks may face transitional behaviors, heterogenous sizes of the collision domains, significant mismatch between transmission rates of control and data messages, prolonged backoff time, and alignment-throughput tradeoff. This paper discussed how the MAC layer functions of existing mmWave standards are not effective in addressing these new challenges. It was argued that the use of new collision-aware hybrid resource allocation, more efficient retransmission policies, collision notification, and multihop communication has the potential to significantly improve the performance of short range mmWave networks.}

\bibliographystyle{IEEEtran}
\bibliography{bibfile}
\end{document}